\documentclass{SciPost}

\hypersetup{
    colorlinks,
    linkcolor={red!50!black},
    citecolor={blue!50!black},
    urlcolor={blue!80!black}
}

\usepackage[bitstream-charter]{mathdesign}
\DeclareSymbolFont{usualmathcal}{OMS}{cmsy}{m}{n}
\DeclareSymbolFontAlphabet{\mathcal}{usualmathcal}

\fancypagestyle{SPstyle}{
\fancyhf{}

}

\begin{document}

\pagestyle{SPstyle}

\begin{center}{\Large \textbf{\color{scipostdeepblue}{
Two faces of Hawking radiation and thin-shell emission: pair-creation vs. tunneling
}}}\end{center}

\begin{center}
\textbf{
Dong-han Yeom$\star$
}
\end{center}

\begin{center}
Department of Physics Education,
\\Pusan National University, Busan 46241, Republic of Korea
\\
Research Center for Dielectric and Advanced Matter Physics,\\
Pusan National University, Busan 46241, Republic of Korea
\\[\baselineskip]
$\star$ \href{mailto:innocent.yeom{}@{}gmail.com}{\small \sf innocent.yeom{}@{}gmail.com}
\end{center}

\definecolor{palegray}{gray}{0.95}
\begin{center}
\colorbox{palegray}{
  \begin{tabular}{rr}
  \begin{minipage}{0.1\textwidth}
  \end{minipage}
  &
  \begin{minipage}{0.75\textwidth}
    \begin{center}
    {\it 4th International Conference on Holography,\\ 
    String Theory and Discrete Approach}\\
    {\it Hanoi, Vietnam, 2020} \\
    \end{center}
  \end{minipage}
\end{tabular}
}
\end{center}

\section*{\color{scipostdeepblue}{Abstract}}
{\bf
We first revisit Hartle and Hawking's path integral derivation of Hawking radiation. In the first point of view, we interpret that a particle-antiparticle pair is created and the negative energy antiparticle falls into the black hole. On the other point of view, a particle inside the horizon, or beyond the Einstein-Rosen bridge, tunnels to outside the horizon, where this computation requires the analytic continuation of the time. These two faces of the Hawking radiation process can be extended to not only particles but also fields. As a concrete example, we study the thin-shell tunneling process; by introducing the antishell as a negative tension shell, we can give the consistent interpretation for two pictures, where one is a tunneling from inside to outside the horizon using instantons, while the other is a shell-antishell pair-creation. This shows that the Euclidean path integral indeed carries vast physical implications not only for perturbative, but also for non-perturbative processes.
}

\vspace{10pt}
\noindent\rule{\textwidth}{1pt}
\tableofcontents
\noindent\rule{\textwidth}{1pt}
\vspace{10pt}

\section{Introduction}

The tension between gravity and quantum mechanics is the fundamental but unresolved problem of modern physics. Although this is a difficult problem, there is one way to approach the nature of quantum gravity. That is to consider quantum effects around a given classical metric, where this is known as semi-classical gravity. This semi-classical approach is not the complete theory, but it can include some useful and important consequences that should be implemented by the consistent quantum theory of gravity. One of such consequences in black hole physics is the emission of particles from the event horizon, or so-called Hawking radiation \cite{Hawking:1974sw}.

The existence of Hawking radiation was already confirmed by various approaches. The most famous explanation is to use the Bogoliubov transformation \cite{Hawking:1974sw}. Then, due to the redshift of the incoming modes around the horizon, the expectation value of the number operator of a given mode at future infinity is non-vanishing, even though the state was chosen to be a vacuum at the past infinity. This approach includes very essential properties of Hawking radiation, but there remain questions. For example, what physically happens at the event horizon?

As a response of the question, one may say that there are several alternative interpretations. First, in the tunneling picture \cite{Hartle:1976tp,Parikh:1999mf,Padmanabhan:2019yyg,Berezin:1997fn}, a particle tunnels from inside to outside the horizon. Second, in the renormalized energy-momentum tensor picture \cite{Davies:1976ei}, a negative energy flux falls into the black hole. Of course, this must be two faces of the same physics. In fact, there is a well-known correspondences between antiparticle and particle; either a particle moves backward in time or an antiparticle moves forward in time.

In this paper, we first revisit the tunneling picture which was developed by Hartle and Hawking \cite{Hartle:1976tp}. The original motivation of this was to explain that a particle tunnels from inside to outside, or equivalently, an antiparticle falls in and a particle comes out from the black hole. In order to evaluate the probability, Hartle and Hawking used the analytic continuation of the time of the inside the horizon. However, we will observe that this is not the unique analytic continuation; one can also use the analytic continuation outside the horizon.

Then, this opens a possibility that the analytic continuation of the time outside the horizon justifies the use of the Euclidean instantons; indeed, Hawking radiation can be interpreted as instantons \cite{Chen:2018aij}. One strong point of the instanton is that one can further extend to non-perturbative processes \cite{Farhi:1989yr}. In this point of view, will the correspondence between two pictures, \textit{the particle-antiparticle pair-creation vs. the tunneling over the event horizon or the Einstein-Rosen bridge}, still be true even in the non-perturbative limit?

Although we cannot provide a complete proof about this, our answer is positive. We will focus on the thin-shell tunneling issue \cite{Sasaki:2014spa} and first construct a tunneling channel which is very similar to the Hartle-Hawking's particle tunneling. From this tunneling instanton, we will reconstruct an analogous process of the particle-antiparticle pair creation; so to speak, a shell-antishell pair creation. We will explain that two pictures give consistent probability interpretations.

This paper is organized as follows. In Sec.~\ref{sec:par}, we first revisit the particle tunneling picture of Hartle and Hawking. We can introduce the analytic continuation of the time of not only inside but also outside the event horizon. We conclude that a particle-antiparticle pair-creation is indeed equivalent to a tunneling over the Einstein-Rosen bridge, or the event horizon. In Sec.~\ref{sec:ext}, we approximately extend this to the field level using the Euclidean analytic continuation and conclude that if there is a positive tension shell, its counter part (antishell) can be described by a negative tension shell, which can be supported by the complexification of the field. In Sec.~\ref{sec:thi}, as a very natural extension, we consider a shell-antishell pair creation process, where we can provide two points of view; one is a shell tunnels from inside to outside, while the other is a negative tension shell falls into the horizon. Finally, in Sec.~\ref{sec:con}, we summarize our results and discuss possible future extensions.

\section{\label{sec:par}Hawking radiation as tunneling: revisit Hartle and Hawking}

In this section, we review the technique introduced by Hartle and Hawking \cite{Hartle:1976tp} and add more comments on possible new interpretations.

\subsection{Formalism}

The tunneling amplitude of a particle with energy $\omega$ from $x = \left(t, \vec{r}\right)$ to $x' = \left(t', \vec{r}'\right)$ (fixing $t' = 0$) is
\begin{eqnarray}
\mathcal{S}\left(\vec{r}', \vec{r}\right) = \int_{-\infty}^{+\infty} dt e^{-i\omega t} K\left(0, \vec{r}'; t, \vec{r}\right),
\end{eqnarray}
where
\begin{eqnarray}
K\left(0, \vec{r}'; t, \vec{r}\right) = -\frac{i}{4\pi^{2}} \frac{1}{s(x,x') \pm i\epsilon}
\end{eqnarray}
is the propagator and $s(x,x')$ is the square of the geodesic distance between $x$ and $x'$.

Note that there is a propagator dependence. If we choose the Feynman propagator, then for $x' > x$, the pole is shifted to the upper direction, while if $x' < x$, the pole is shifted to the lower direction. On the other hand, by choosing advanced or retarded propagator, one can choose the location of the poles by different ways.

Since the propagator only depends on $t' - t$ and $s(x',x)$ is the square of the geodesic distance, without loss of generality, we obtain
\begin{eqnarray}
K\left(0, \vec{r}'; t, \vec{r}\right) = K\left(t, \vec{r}; 0, \vec{r}'\right) = K\left(-t, \vec{r}'; 0, \vec{r}\right).
\end{eqnarray}
By using the second term, in order to evaluate the amplitude, we interpret that we fix a point outside the black hole $x'$ and integrate over $t$ by sliding the point inside the black hole $x$, which is the same as Hartle and Hawking. On the other hand, if we use the third term, we interpret that we fix a point inside the black hole $x$ and integrate over $t$ by sliding the point outside the black hole $x'$. Then we will get the same result but obtain a different interpretation.

\subsection{First point of view: particle-antiparticle pair creation}

We recall the integration
\begin{eqnarray}
\mathcal{S}\left(\vec{r}', \vec{r}\right) = \int_{-\infty}^{+\infty} dt e^{-i\omega t} K\left(-t, \vec{r}'; 0, \vec{r}\right).
\end{eqnarray}
If we redefine $t = -T$, then this integration becomes
\begin{eqnarray}
\mathcal{S}\left(\vec{r}', \vec{r}\right) = \int_{-\infty}^{+\infty} dT e^{-i (-\omega) T} K\left(0, \vec{r}; T, \vec{r}'\right).
\end{eqnarray}
Now we can interpret that, in this new time coordinates $T$, this is equivalent to see that a particle of negative energy $-\omega$ propagates from outside to inside the horizon. This means that there is a particle-antiparticle pair-creation outside the horizon; and the antiparticle with negative energy falls into the black hole.

\subsection{Second point of view: tunneling from inside to outside the horizon}

\begin{figure}
\begin{center}
\includegraphics[scale=0.7]{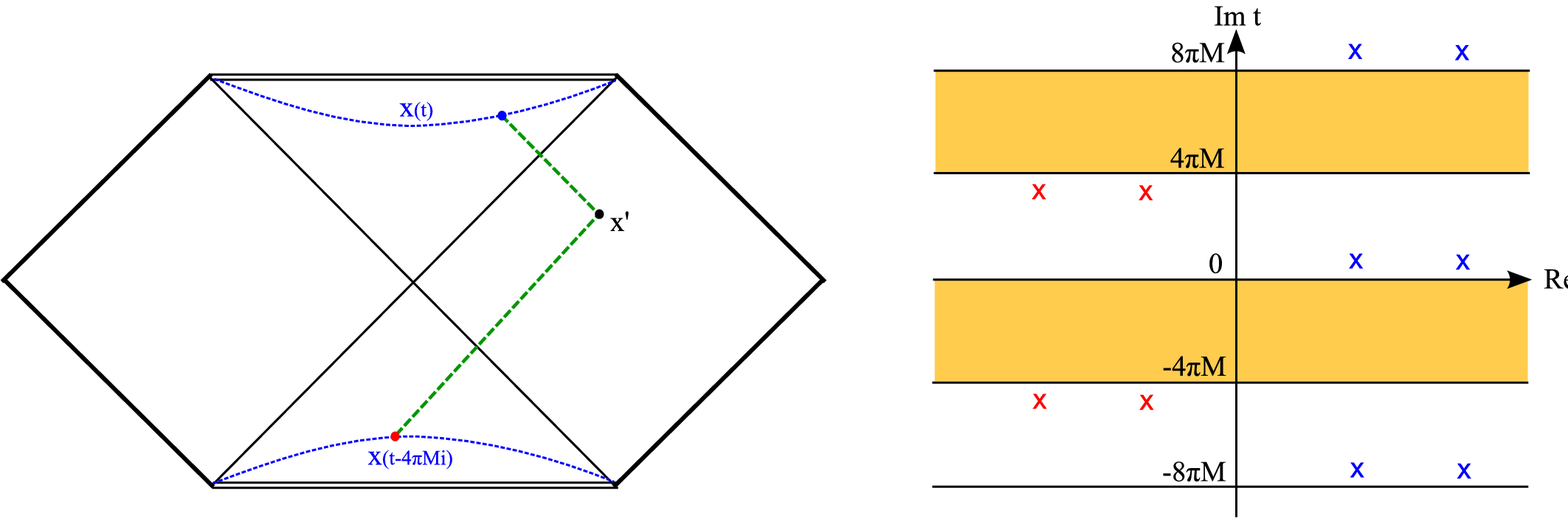}
\includegraphics[scale=0.7]{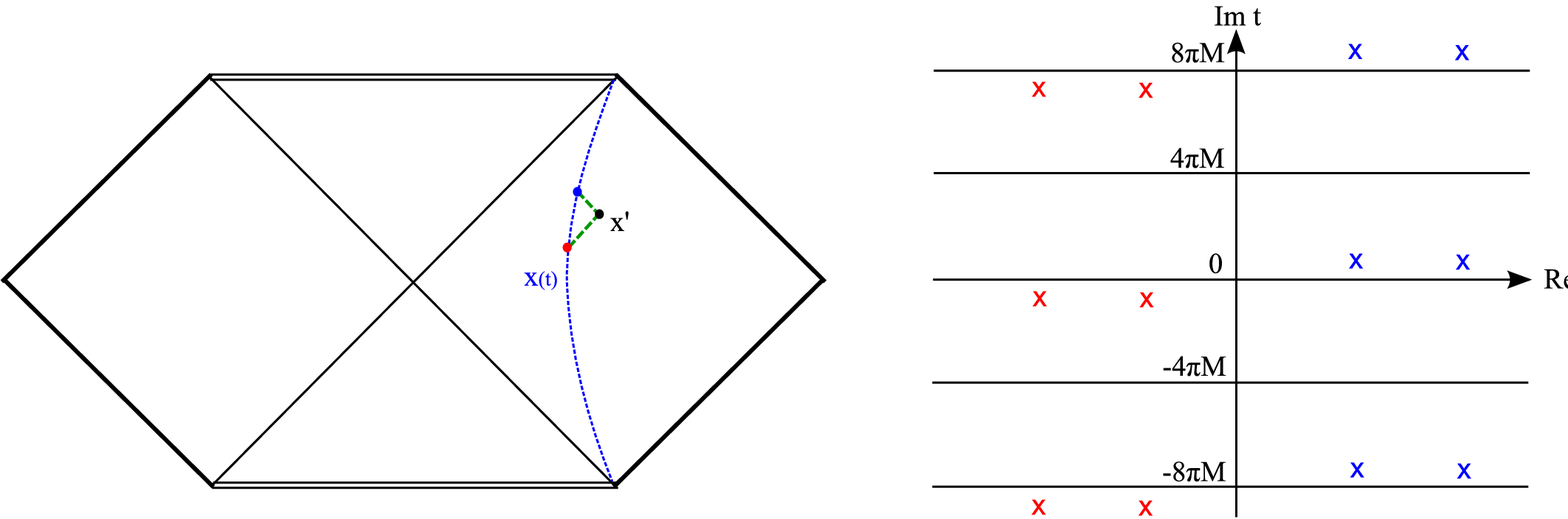}
\caption{\label{fig:pic}Case 1 (upper) and Case 2 (lower). For Case 1 and 2, $x'$ is fixed and $x$ is sliding over time $t$. The green dashed curves show two possible points which can be connected by null geodesics. Each null geodesic is denoted by a pole in the complex time plain and the colors (red and blue dots) corresponds x-marks in the right diagram. For each color, there are more than one x-mark, because longer geodesics can be obtained by considering $\theta$ and $\varphi$ angles. The poles are shifted by $\pm i\epsilon$ due to the Feynman propagator.}
\end{center}
\end{figure}

\begin{figure}
\begin{center}
\includegraphics[scale=0.7]{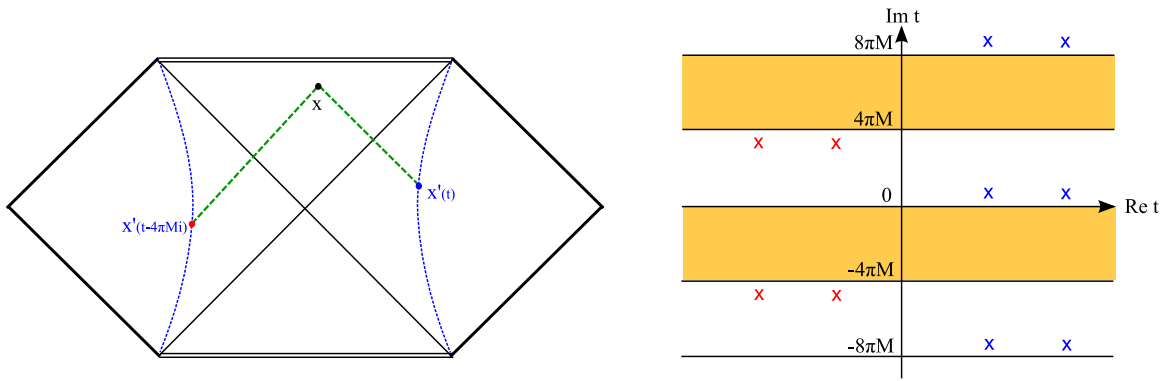}
\caption{\label{fig:pic3}Case 3. For Case 3, $x$ is fixed and $x'$ is sliding. Note that for Case 1, 
$x(t-4\pi M i)$ is the causal past of $x'$, while for Case 3, $x$ is always causal future of $x'$. Hence, in order to have the same analytic structure with Case 1, we need to introduce the advanced propagator for the poles at $x'(t-4\pi M i)$.}
\end{center}
\end{figure}

Now we see the analytic structure of $K\left(0, \vec{r}'; t, \vec{r}\right)$ as a function of $t$ by fixing $\vec{r}$ and $\vec{r}'$. The important thing to check is the pole structure, where the pole happens when $x$ and $x'$ can be connected by null geodesics.

We will see the pole structures for three different cases.
\begin{itemize}
\item[-- 1.] Choice of $K\left(t, \vec{r}; 0, \vec{r}'\right)$, while $\vec{r}$ is inside the black hole and $\vec{r}'$ is outside the black hole (upper of Fig.~\ref{fig:pic}): Following Hartle-Hawking, there appears poles in $t$ (inside the black hole region) as well as $t - 4\pi Mi$ (inside the white hole region) lines. There are two kinds of poles, where one is $x' > x$ (at $\mathrm{Im} t = 0$) and the other one is $x' < x$ (at $\mathrm{Im} t = -4 \pi M i$). Hence, by introducing the Feynman propagator, one can shift the poles to the upper direction for the former and the lower direction for the latter. Then, due to the analyticity, one can shift the time integration from $t$ to $t - 4\pi M i$.
\item[-- 2.] Choice of $K\left(t, \vec{r}; 0, \vec{r}'\right)$, while both of $\vec{r}$ and $\vec{r}'$ are outside the black hole (lower of Fig.~\ref{fig:pic}): Again, as described by Hartle-Hawking, there appears poles only in the $t$ line. There are two kinds of poles, where one is $x' > x$ and the other one is $x' < x$. Hence, by introducing the Feynman propagator, the half of the poles is shifted to the upper direction and the other half of the poles is shifted to the lower direction. Then, in the end, there is no good way to analytically continue to the Euclidean time.
\item[-- 3.] Choice of $K\left(-t, \vec{r}'; 0, \vec{r}\right)$, while $\vec{r}$ is inside the black hole and $\vec{r}'$ is outside the black hole (Fig.~\ref{fig:pic3}): Now $\vec{r}$ can be connected via null geodesics to $\vec{r}'$ by two ways: one is in the $t$ side (right side of the Penrose diagram), while the other one is in the $t - 4 \pi M i$ side (left side of the Penrose diagram). Note that both of them satisfy $x' > x$. Therefore, in order to obtain the same analytic structure, we need to introduce the retarded propagator for the former poles and the advanced propagator for the latter poles.
\end{itemize}

\begin{figure}
\begin{center}
\includegraphics[scale=0.6]{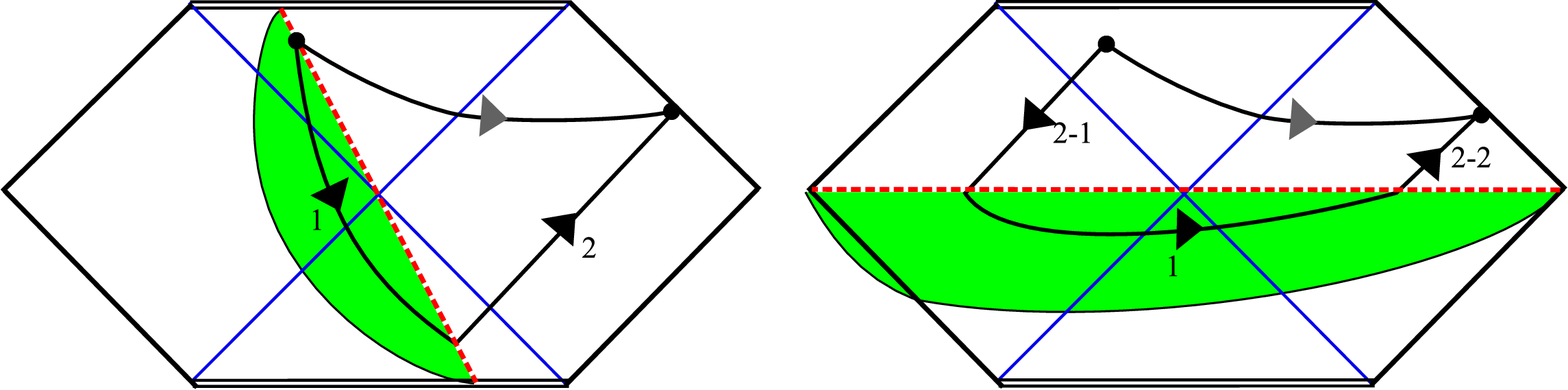}
\caption{\label{fig:cycles}Comparison with Case 1 (left) and Case 3 (right). The contour 2 of left should be identified to the contour 2-1 and 2-2 of right. This requires the advanced propagator for 2-1.}
\end{center}
\end{figure}

Therefore, the choice of the propagator for Case 3 means that we choose such a propagator in order to make the contour integration proportional to the absorption probability (Fig.~\ref{fig:cycles}). In terms of the particle trajectory, for Case 1 (left), one can approximately understand that the tunneling rate is calculated by two parts, where one is over the Euclidean manifold (trajectory 1, inside the horizon) and the other is from inside the white hole to the future infinity (trajectory 2), while the latter is the same as the absorption probability. The same process is described by Case 3 (right), and hence, the absorption probability is described by the trajectories 2-1 and 2-2. This means that in order to interpret this contribution as the absorption probability, the causality of 2-1 should be regarded as opposite to the usual way.

Finally, for Case 1 and Case 3, due to the analytic structure, we can shift the time integration from $t$ to $t - 4\pi M i$. Then we obtain the following integration
\begin{eqnarray}
\mathcal{S}\left(\vec{r}', \vec{r}\right) = e^{-4\pi M \omega} \int_{-\infty}^{+\infty} dt e^{-i\omega t} K\left(0, \vec{r}'; t - 4\pi M i, \vec{r}\right).
\end{eqnarray}
After squaring both sides, we obtain
\begin{eqnarray}
(\mathrm{emission\;probability}) = e^{-8\pi M \omega} \times (\mathrm{absorption\; probability}),
\end{eqnarray}
that means the Hawking temperature is $T = 1/8\pi M$.

\section{\label{sec:ext}Extension to fields and shells}

Now we understand that two points of view are equivalent, in other words, the in-coming negative energy particle is equivalent to a positive energy particle that comes out through the Einstein-Rosen bridge. This equivalence can be generalized to field level descriptions if we interpret a field as a bunch of particles.

Let us consider a propagator between an initial hypersurface defined on past infinity $(h^{\mathrm{in}}_{ab}, \phi^{\mathrm{in}})$ to a final hypersurface defined on the future infinity $(h^{\mathrm{out}}_{ab}, \phi^{\mathrm{out}})$:
\begin{eqnarray}
\Psi \left[ h^{\mathrm{out}}_{ab}, \phi^{\mathrm{out}}; h^{\mathrm{in}}_{ab}, \phi^{\mathrm{in}} \right] = \int \mathcal{D}g_{\mu\nu} \mathcal{D}\phi \;\; e^{i S[g_{\mu\nu},\phi]},
\end{eqnarray}
where we sum over all metric $g_{\mu\nu}$ and matter field $\phi$ those connect from $(h^{\mathrm{in}}_{ab}, \phi^{\mathrm{in}})$ to $(h^{\mathrm{out}}_{ab}, \phi^{\mathrm{out}})$. For the ground state, we may introduce the Euclidean analytic continuation and introduce the Euclidean path integral \cite{Hartle:1983ai} such as
\begin{eqnarray}
\Psi_{0} \left[ h^{\mathrm{out}}_{ab}, \phi^{\mathrm{out}}; h^{\mathrm{in}}_{ab}, \phi^{\mathrm{in}} \right] = \int \mathcal{D}g_{\mu\nu} \mathcal{D}\phi \;\; e^{- S_{\mathrm{E}}[g_{\mu\nu},\phi]}.
\end{eqnarray}
This Euclidean path integral can be further approximated by two assumptions \cite{Chen:2018aij}:
\begin{itemize}
\item[--] First, we can assume a kind of symmetry and restrict to the mini-superspace, e.g., the spherical symmetry.
\item[--] Second, we can approximate the path-integral by using the steepest-descent approximation, so-called sum-over instantons:
\begin{eqnarray}
\Psi_{0} \left[ h^{\mathrm{out}}_{ab}, \phi^{\mathrm{out}}; h^{\mathrm{in}}_{ab}, \phi^{\mathrm{in}} \right] \simeq \sum_{\mathrm{on-shell}} e^{- S_{\mathrm{E}}^{\mathrm{on-shell}}[g_{\mu\nu},\phi]}.
\end{eqnarray}
\end{itemize}

One interesting point is that because of the analytic continuation, every fields should be complexified \cite{Hartle:2007gi}. On the other hand, we need to impose the reality condition for the initial hypersurface as well as the final hypersurface. This means that for the intermediate geometry between the initial and the final surfaces, it is allowed to introduce imaginary valued fields, as long as the imaginary field value does not reach the future infinity.

In this paper, we will restrict the complexification of only matter fields for simplicity. Then what are the effective roles of the imaginary part, especially, of the scalar field? Let us focus on the following points. For a (classical) scalar field, for a given time slice, we can use the Fourier transform and split each different momentum modes. 

The scalar field of each mode will be proportional to $\phi_{\omega} \propto a(\omega) e^{-i\omega t}$. If the internal proper time $t$ of the scalar field is following the backward direction, then it is equivalent to change $\omega \rightarrow - \omega$; the energy becomes negative. Or, equivalently, if we introduce overall imaginary factor $i = \sqrt{-1}$ to the scalar field, i.e., $\phi_{\omega} \rightarrow i\phi_{\omega}$, then it is also equivalent to consider a negative kinetic energy (hence, an effective ghost field \cite{Chen:2015ria}), because the effective number of each mode becomes negative $|a(\omega)|^{2} \rightarrow -|a(\omega)|^{2}$. To summarize, for a given mode, the follows are equivalent:
\begin{itemize}
\item[--] A real scalar field with positive energy moves \textit{backward} in time,
\item[--] A real scalar field with \textit{negative energy} moves forward in time,
\item[--] An \textit{imaginary} scalar field with positive energy moves forward in time. 
\end{itemize}
Because of these relations, it is justifiable to introduce a negative tension shell (as long as it does not reach to past or future infinity), that is motivated from the complexified scalar field. For simplicity, we interpret that these imaginary (or negative tension) shell as a positive tension shell that moves the opposite way of the coordinate time.

\section{\label{sec:thi}Thin-shell instantons revisited}

In this section, we apply for the results of the previous sections. First, for the Hawking radiation, we have two points of view for the same phenomenon: (1) a particle-antiparticle pair-creation and (2) a particle tunneling from inside to outside the horizon, or a particle tunnels through the Einstein-Rosen bridge, where the latter can be described by Euclidean approaches. Second, if we extend this to the field level, then the notion of the antiparticle can be well matched to the analytically continued imaginary part of the field. In the thin-shell case, the imaginary field will effectively give a negative tension, because its kinetic term is negative.

Now we have the following question. Will the thin-shell tunneling process (which is usually described by the Euclidean instantons \cite{Farhi:1989yr,Gregory:2013hja}) be equivalently described by the shell-antishell (or, positive-tension-shell-negative-tension-shell) pair-creation process? Will these two pictures be consistent? In this section, we will answer for these questions.

\subsection{Thermal single-shell: junction equation and causal structures}

We first follow rather a canonical approach: a tunneling process of a positive tension shell. We consider a spacetime with the spherical symmetric metric ansatz
\begin{eqnarray}
\label{eq:metric}
ds_{\pm}^{2}= - f_{\pm}(R) dT^{2} + \frac{1}{f_{\pm}(R)} dR^{2} + R^{2} d\Omega^{2},
\end{eqnarray}
where we prepare a thin-shell that locates at $r$: outside the shell is $r < R$ (denoted by $+$) and inside the shell is $R < r$ (denoted by $-$). The thin-shell will satisfy the metric
\begin{eqnarray}
ds^{2} = - dt^{2} + r^{2}(t) d\Omega^{2}.
\end{eqnarray}
We impose the metric ansatz for outside and inside the shell
\begin{eqnarray}
f_{\pm}(R) = 1 - \frac{2M_{\pm}}{R} - \frac{R^{2}}{\ell_{\pm}^{2}}.
\end{eqnarray}
Here, $M_{+}$ and $M_{-}$ are the mass parameters of each region and
\begin{eqnarray}
\ell^{2}_{\pm} = \frac{3}{8\pi V_{\pm}}
\end{eqnarray}
is the parameter due to the vacuum energy $V_{\pm}$. We assume $M_{+} > M_{-}$ (for the opposite case, see \cite{Kang:2017hpz}).

The equation of motion of the thin-shell is determined by the junction equation \cite{Israel:1966rt}:
\begin{eqnarray}\label{eq:junc}
\epsilon_{-} \sqrt{\dot{r}^{2}+f_{-}(r)} - \epsilon_{+} \sqrt{\dot{r}^{2}+f_{+}(r)} = 4\pi r \sigma.
\end{eqnarray}
Here, $\epsilon_{\pm} = \pm 1$, where these $\epsilon$ parameters denotes the outward normal direction of the shell. $\sigma$ is the tension parameter and we choose positive value so that this satisfies the null energy condition.

After simple calculations, we can reduce the junction equation by a simpler formula \cite{Blau:1986cw}:
\begin{eqnarray}\label{eq:form}
\dot{r}^{2} + V(r) &=& 0,\\\label{eq:form2}
V(r) &=& f_{+}(r)- \frac{\left(f_{-}(r)-f_{+}(r)-16\pi^{2} \sigma^{2} r^{2}\right)^{2}}{64 \pi^{2} \sigma^{2} r^{2}}.
\end{eqnarray}
Now we consider the nucleation of thin-shells. Such a nucleation can be explained by an instanton. For simplicity, we first consider the case of the thermal excitations \cite{Garriga:2004nm,Chen:2017suz}, where this requires the conditions $V(r_{s}) = V'(r_{s}) = 0$ for a given $r_{s}$; later, we will extend to a generic instanton. The detailed condition for the thermal excitation condition is described in the Appendix of \cite{Yeom:2016qec}.

After we classify the classical trajectories, we determine signs of $\epsilon$ parameters by comparing extrinsic curvatures:
\begin{eqnarray}\label{eq:ec1}
\beta_{\pm}(r) \equiv \frac{f_{-}(r)-f_{+}(r)\mp 16\pi^{2} \sigma^{2} r^{2}}{8 \pi \sigma r} = \epsilon_{\pm} \sqrt{\dot{r}^{2}+f_{\pm}(r)}.
\end{eqnarray}
If we choose $\ell_{+} = \infty$, $\ell_{-}^{-2} < 0$, and $- \ell_{-}^{-2} - 16 \pi^{2} \sigma^{2} > 0$, then the only allowed solution is $\epsilon_{\pm} = +1$ for all regions.

In the Euclidean signatures, the shell is in a stable local minimum and will maintain a constant radius. Therefore, the causal structure is easy to explain. On the other hand, it is on a unstable local maximum in the Lorentzian signatures. Therefore, after a nucleation, it is reasonable to think that the shell either collapses or expands \cite{Chen:2017suz}.

\subsection{Thermal double-shell: junction equations and causal structures}

Now we ask whether the same process can be described by a shell-antishell process. In order to do this, we need to introduce not only a positive tension shell, but also a negative tension shell. Due to the symmetry, it is not difficult to extend the thin-shell formalism.

We consider spacetime with the metric,
\begin{eqnarray}
ds_{+,0,-}^{2} = - f_{+,0,-}(R) dT^{2} + \frac{1}{f_{+,0,-}(R)} dR^{2} + R^{2} d\Omega^{2}.
\end{eqnarray}
We prepare two thin-shells where one is the outer shell $r_{2}$ and the other is the inner shell $r_{1}$ ($r_{1} \leq r_{2}$): outside the outer shell is $r_{2} < R$ (denoted by $+$), inside the inner shell is $R < r_{1}$ (denoted by $-$), and the intermediate region is $r_{1} < R < r_{2}$ (denoted by $0$).

Thin-shells will follow the metric
\begin{eqnarray}
ds^{2} = - dt^{2} + r_{1,2}^{2}(t) d\Omega^{2}.
\end{eqnarray}
We impose the metric ansatz for outside and inside the shell
\begin{eqnarray}
f_{+,0,-}(R) = 1 - \frac{2M_{+,0,-}}{R} - \frac{R^{2}}{\ell_{+,0,-}^{2}}.
\end{eqnarray}
The equations of motion of thin-shells are as follows:
\begin{eqnarray}
\epsilon^{(2)}_{0} \sqrt{\dot{r}_{2}^{2}+f_{0}(r_{2})} - \epsilon^{(2)}_{+} \sqrt{\dot{r}_{2}^{2}+f_{+}(r_{2})} &=& 4\pi r_{2} \sigma_{2},\\
\epsilon^{(1)}_{-} \sqrt{\dot{r}_{1}^{2}+f_{-}(r_{1})} - \epsilon^{(1)}_{0} \sqrt{\dot{r}_{1}^{2}+f_{0}(r_{1})} &=& 4\pi r_{1} \sigma_{1}.
\end{eqnarray}
Here, $\epsilon^{(1,2)}_{+,0,-} = \pm 1$ to denote the outward normal directions. $\sigma_{1,2}$ are tension parameters.

If we assume $M_{+}=M_{-}=M$, $\sigma_{2} = - \sigma_{1} = \sigma$, $\ell_{+} = \ell_{-} = \ell$, and $\Delta_{\pm} \equiv M \pm M_{0}$, then
\begin{eqnarray}
\dot{r}_{1,2}^{2} + V_{1,2}(r_{1,2}) = 0,
\end{eqnarray}
where
\begin{equation}
V_{2}(r_{2}) = 1 - \left(\frac{\rho_{+}^{2}}{64 \pi^{2} \sigma^{2}} + \frac{\ell_{0}^{-2} + \ell_{+}^{-2}}{2} + 4\pi^{2} \sigma^{2} \right) r_{2}^{2} - \left( 1 + \frac{\rho_{+}}{16 \pi^{2} \sigma^{2}} \frac{\Delta_{-}}{\Delta_{+}} \right) \frac{\Delta_{+}}{r_{2}} - \frac{\Delta_{-}^{2}}{16\pi^{2}\sigma^{2} r_{2}^{4}},
\end{equation}
\begin{equation}
V_{1}(r_{1}) = 1 - \left(\frac{\rho_{+}^{2}}{64 \pi^{2} \sigma^{2}} + \frac{\ell_{+}^{-2} + \ell_{0}^{-2}}{2} + 4\pi^{2} \sigma^{2} \right) r_{1}^{2} - \left( 1 + \frac{\rho_{+}}{16 \pi^{2} \sigma^{2}} \frac{\Delta_{-}}{\Delta_{+}} \right) \frac{\Delta_{+}}{r_{1}} - \frac{\Delta_{-}^{2}}{16\pi^{2}\sigma^{2} r_{1}^{4}},
\end{equation}
and hence the effective potential coincide.

After we classify the classical trajectories, we determine signs of $\epsilon$ parameters by comparing extrinsic curvatures:
\begin{eqnarray}
\beta^{(2)}_{+}(r_{2}) &\equiv& \frac{f_{0}(r_{2})-f_{+}(r_{2})-16\pi^{2} \sigma_{2}^{2} r_{2}^{2}}{8 \pi \sigma_{2} r_{2}} = \epsilon^{(2)}_{+} \sqrt{\dot{r}_{2}^{2}+f_{+}(r_{2})},\\
\beta^{(2)}_{0}(r_{2}) &\equiv& \frac{f_{0}(r_{2})-f_{+}(r_{2})+16\pi^{2} \sigma_{2}^{2} r_{2}^{2}}{8 \pi \sigma_{2} r_{2}} = \epsilon^{(2)}_{0} \sqrt{\dot{r}_{2}^{2}+f_{0}(r_{2})},\\
\beta^{(1)}_{0}(r_{1}) &\equiv& \frac{f_{-}(r_{1})-f_{0}(r_{1})-16\pi^{2} \sigma_{1}^{2} r_{1}^{2}}{8 \pi \sigma_{1} r_{1}} = \epsilon^{(1)}_{0} \sqrt{\dot{r}_{1}^{2}+f_{0}(r_{1})},\\
\beta^{(1)}_{-}(r_{1}) &\equiv& \frac{f_{-}(r_{1})-f_{0}(r_{1})+16\pi^{2} \sigma_{1}^{2} r_{1}^{2}}{8 \pi \sigma_{1} r_{1}} = \epsilon^{(1)}_{-} \sqrt{\dot{r}_{1}^{2}+f_{-}(r_{1})}.
\end{eqnarray}

Let us define $M_{+} = M_{-} = M$ and $\sigma_{2} = - \sigma_{1} = \sigma$, so that the outer shell is a real shell and the inner shell is an imaginary shell. We assume that $\ell = \infty$, $\ell_{0}^{-2} < 0$, and $|4 \pi \sigma \ell_{0}|^{-2} > 1$. Then, the equation of motion for each shell will follow the same potential that was discussed in the previous section. In addition, since the tension of the inner shell is negative, the signs of the extrinsic curvatures should be always positive for both of inner and outer shells. Therefore, from the point of crossing radius $r_{s}$, the allowed solutions are only two types those are located right patch of the Penrose diagram: branching and emerging solutions.

We consider an Euclidean analytic continuation when the effective potential satisfies the matching condition $V'(r_{s}) = V(r_{s}) = 0$. In this limit, if we fold two shells at $r = r_{s}$, then $\dot{r} = 0$; in Euclidean signatures this is a stable local minimum. Also, since one positive and one negative shells are folded, effectively the Euclidean Schwarzschild manifold with mass $M$ is the entire Euclidean solution.

After analytic continuation to the Lorentzian time, the order of $+$ part and $-$ part are changed for the left side of the causal patch. So, for the right part, we can continue to a trivial branching shell solution. On the other hand, for the left part, the directions of outward normal vectors are flipped. For these flipped outward normal vectors, the only possible corresponding solution is the rotated result of the emerging shells.

\subsection{Probability: equivalence to the single-shell interpretation}

For the thermal instantons, the velocity of the shell is zero on the Euclidean manifold and maintains a constant radius $r_{0}$. Since there is no dynamics of the shell on the Euclidean manifold, the only contribution of the Euclidean action comes from the regularization of the cusp of the horizon. We can calculate the probability of a single shell based on the Euclidean path integral formalism \cite{Gregory:2013hja}, where the result is $P \sim e^{-2 B}$ with
\begin{eqnarray}
2B = 2 \left[S_{\mathrm{E}}(\mathrm{solution}) - S_{\mathrm{E}}(\mathrm{background}) \right] = - \frac{\Delta \mathcal{A}}{4},
\end{eqnarray}
where $\mathcal{A}$ is the area of the horizon. For the evaporating case, $\Delta \mathcal{A} < 0$ and hence $2B$ is positive definite.

On the other hand, if we interpret the shell-antishell pair-creation process and regard that the negative energy falls into the black hole, the areal entropy is then changed while the energy is conserved. Then after a sufficient Lorentzian time, we compare two solutions, where one is the pure Schwarzschild solution with mass $M_{+}$ and the other is the Schwarzschild-AdS solution with mass $M_{-}$ and the cosmological constant $V_{-}$. We can evaluate the Euclidean action difference between the two solutions, because the shell dynamics will not contribute to the probability during the Lorentzian time. As a result, the action difference between two solutions will be the same as the areal entropy difference. As a simple check, we can evaluate following the thermodynamic way. We can estimate the Helmholtz free energy difference as follows \cite{Chen:2017suz}:
\begin{eqnarray}
2B = \frac{\Delta \mathcal{F}}{T} = - \frac{\Delta \mathcal{A}}{4},
\end{eqnarray}
where $\mathcal{F} = E - ST$ is the Helmholtz free energy, $E$ is the energy, $S = \mathcal{A}/4$ is the areal entropy, and the last equation is obtained since the energy is conserved ($\Delta E = 0$).

Therefore, we obtain the consistent and equivalent probabilities from two pictures: from the thin-shell tunneling picture, the probability is decided by the Euclidean action integral, while the shell-antishell pair-creation picture, the probability can be interpreted by the thermodynamic way.

\subsection{Beyond the thermal shell limit}

If one goes beyond the thermal shell, the shell will have dynamics in the Euclidean domain. Usually, one can interpret this process such that this is a tunneling from a small radius (say, $r_{1}$) to a larger radius (say, $r_{2}$). Or, alternatively, one can interpret that two shells are created at the same time, where one is left side and the other is right side of the Einstein-Rosen bridge. The difference of two interpretations is whether there is a contribution of the areal entropy dependence, where the former has no such a contribution while the latter must have the term from the regularization of the cusp of the Euclidean manifold.

Now what happens if we apply this for the double-shell model? We will interpret that the negative tension shell will collapse from the smaller radius, while the positive tension shell will expand from the larger radius in the Lorentzian domain. Then, in the Euclidean domain, two shells must meet each other. Note that the Euclidean action contribution becomes \cite{Gregory:2013hja}
\begin{eqnarray}
2B = \frac{\mathcal{A}_{\mathrm{i}} - \mathcal{A}_{\mathrm{f}}}{4} + 2 \int_{r_{1}}^{r_{2}} dr r \left| \cos^{-1} \left( \frac{f_{+} + f_{-} - 16 \pi^{2} \sigma^{2} r^{2}}{2\sqrt{f_{+}f_{-}}} \right) \right|,
\end{eqnarray}
where the last term of the right hand side is due to the dynamics of the shell. Note that this is symmetric up to the change of $f_{\pm} \rightarrow f_{\mp}$ and $\sigma \rightarrow -\sigma$. Therefore, if the two shells meet at $r=r_{c}$, then the positive tension shell integration from $r_{1}$ to $r_{2}$ is the same as the negative tension shell integration from $r_{1}$ to $r_{c}$ plus the positive tension shell integration from $r_{c}$ to $r_{2}$. If we interpret the first term using the thermodynamic way (as the negative tension shell collapses, the area decreases and hence the entropy changes), then we will eventually obtain the same probabilistic interpretation \cite{Chen:2020nag}.

One potential assumption is that there is no singular contribution when two shells meet each other. This is beyond the scope of the thin-shell approximation. We leave this detailed clarification for a future research topic.

\section{\label{sec:con}Conclusion: toward generic complex-valued instantons}

In this paper, we first revisited the particle tunneling from a black hole using the Hartle-Hawking's tunneling picture. One can interpret this process as a pair-creation of particle and antiparticle, where the antiparticle moves from outside to inside the horizon by carrying negative energy. On the other hand, in order to evaluate the tunneling probability, one needs to introduce the Euclidean analytic continuation of the time; original Hartle and Hawking introduced the analytic continuation inside the horizon, but equivalently, we can introduce the analytic continuation outside the horizon. Therefore, we can interpret that the pair-creation is equivalent to a particle tunneling from inside to outside the horizon, where initially the particle is located beyond the Einstein-Rosen bridge and connected by the Euclidean manifold.

We can ask whether this picture can be extended to more generic contexts. Due to the analytic continuation of the time, in principle, all fields can be analytically continued to complex-valued functions. The imaginary part of the matter field will have a negative kinetic energy, and hence, it can be interpreted by a flux of negative energy antiparticles. If this is accumulated, in the thin-shell approximation, one may describe the antishell as a negative tension shell.

In this paper, we further extended this idea. We could describe the dynamics of the shell by using the junction equation. In addition, we can evaluate the tunneling probability following the Euclidean path integral approach. The same thing can be done by introducing two shells, where one has the positive tension (shell) and the other has the negative tension (antishell). We showed that we obtain the same probability from the two interpretations, not only for thermal shells, but also for generic processes.

Therefore, the correspondence between the pair-creation and the instanton tunneling look very natural and fundamental, not only for the perturbative level (Hawking radiation) but also for the non-perturbative level. This strongly indicate that the Euclidean path integral indeed include very important and essential nature of the quantum gravity. The construction of a shell-antishell is not difficult to imagine at once we have a matter field. One benefit of the shell-antishell interpretation is that the understanding of the causal structure is rather simpler, because we do not need to assume the geometry beyond the Einstein-Rosen bridge. This will allow to understand the physical causal structure of an evaporating black hole with non-perturbative or non-adiabatic processes, where this will do a very important role in order to understand the information loss problem, for example, to show the possibility of a naked black hole firewall \cite{Hwang:2012nn} or to make the Einstein-Rosen bridge traversable \cite{Chen:2016nvj}. In addition, the contributions from the trivial geometries without a singularity might be dominated in the late time \cite{Bouhmadi-Lopez:2019kkt}. This will shed some lights to the answer for the information loss paradox.

Of course, there are several issues that must be clarified. For example, we relied on the thin-shell approximation, but we need to check whether the correspondence can be consistent even beyond the thin-shell approximation, or not. Also, if two shells are collided, can there be any new effects? We leave these topics for future investigations.

\section*{Acknowledgment}

The author would like to thank Pisin Chen and Misao Sasaki. DY was supported by the National Research Foundation of Korea (Grant No.: 2018R1D1A1B07049126).

\end{document}